# Enhancing Interconnect Reliability and Performance by Converting Tantalum to 2D Layered Tantalum Sulfide at Low Temperature

Chun-Li Lo[1,2], Massimo Catalano[3,4], Ava Khosravi[3], Wanying Ge[5], Yujin Ji[5], Dmitry Y. Zemlyanov[2], Luhua Wang[3], Rafik Addou[3], Yuanyue Liu[5], Robert M. Wallace[3], Moon J. Kim[3], and Zhihong Chen[1,2*]

[1] School of Electrical and Computer Engineering, Purdue University, West Lafayette, IN 47907, USA
[2] Birck Nanotechnology Center, Purdue University, West Lafayette, IN 47907, USA
[3] Materials Science and Engineering Department, The University of Texas at Dallas, 800 West Campbell Road, Richardson, TX 75080, USA
[4] Institute for Microelectronics and Microsystems, National Council for Research (IMM-CNR), Via Monteroni, ed. A3, 73100 Lecce, Italy
[5] Texas Materials Institute and Department of Mechanical Engineering, The University of Texas at Austin, Austin, TX 78712, USA

**Abstract**

The interconnect half-pitch size will reach ~20 nm in the coming sub-5 nm technology node. Meanwhile, the TaN/Ta (barrier/liner) bilayer stack has to be > 4 nm to ensure acceptable liner and diffusion barrier properties. Since TaN/Ta occupy a significant portion of the interconnect cross-section and they are much more resistive than Cu, the effective conductance of an ultra-scaled interconnect will be compromised by the thick bilayer. Therefore, two dimensional (2D) layered materials have been explored as diffusion barrier alternatives. However, many of the proposed 2D barriers are prepared at too high temperatures to be compatible with the back-end-of-line (BEOL) technology. In addition, as important as the diffusion barrier properties, the liner properties of 2D materials must be evaluated, which has not yet been pursued. Here, a 2D layered tantalum sulfide ($TaS_x$) with ~1.5 nm thickness is developed to replace the conventional TaN/Ta bilayer. The $TaS_x$ ultra-thin film is industry-friendly, BEOL-compatible, and can be directly prepared on dielectrics. Our results show superior barrier/liner properties of $TaS_x$ compared to the TaN/Ta bilayer. This single-stack material, serving as both a liner and a barrier, will enable continued scaling of interconnects beyond 5 nm node.

*E-mail: zhchen@purdue.edu

**Introduction**

Copper (Cu) has been used as the main conductor in interconnects due to its low resistivity. However, because of its high diffusivity, diffusion barriers must be incorporated to surround Cu wires. Otherwise, Cu ions/atoms will drift/diffuse through the inter-metal dielectric (IMD) that separates two distinct interconnects, resulting in circuit shorting. Conventionally, tantalum nitride (TaN) has been adopted as the diffusion barrier owing to its superior capability of blocking Cu diffusion[1,2]. However, the adhesion of Cu to TaN is not ideal. To address this issue, tantalum (Ta) has been integrated in-between TaN and Cu to improve the adhesion. The optimized stack consists of a TaN layer deposited on low-*k* dielectrics, followed by a Ta layer before the Cu deposition, as depicted in Fig. 1a. This trench structure is also known as the "damascene structure." The barrier/liner (TaN/Ta) bilayer has been demonstrated to fulfill several requirements of the interconnect technology[1].

With the advent of sub-5 nm technology node, the interconnect half-pitch size will reach ~20 nm and below[3]. Meanwhile, the thickness of the conventional TaN/Ta bilayer cannot be reduced below ~4-5 nm just to maintain its capability of blocking Cu diffusion into the IMDs. However, the TaN/Ta bilayer is much more resistive than Cu. In ultra-scaled Cu interconnects, TaN/Ta will occupy a large portion of the cross-section area, which tremendously increases the line resistance, as shown in Fig. 1b. The main challenge of scaling conventional TaN/Ta lies in their three-dimensional (3D) nature. Ultrathin and continuous TaN/Ta stack will be difficult to achieved by the commonly-used sputtering method. Although thinner TaN/Ta films prepared by atomic-layer-deposition (ALD) have been investigated, the diffusion barrier property was known to be compromised[4]. Given their atomically thin nature, two-dimensional (2D) layered materials have been considered as diffusion barrier alternatives[5–13] . Among them, graphene[5–10], molybdenum disulfide (MoS$_2$)[11,13], and hexagonal boron nitride (h-BN)[11] have been demonstrated to have desirable diffusion barrier properties even with sub-nm thicknesses. Although seldomly discussed, it has been discovered experimentally that some 2D materials can suppress inelastic scattering in Cu wires and hence reduce Cu resistivity[12,14,15], which hints at the possibility that 2D materials may have preferable liner properties. The advantages are summarized in Fig. 1c. Nevertheless, most 2D materials are prepared by thermal chemical vapor deposition (CVD) at high-temperature (> 800 °C)[5–11], which is way above the back-end-of-line (BEOL) compatible temperature (≤ 400 °C). Moreover, the transfer process from metal substrates to dielectrics required for some 2D materials[5–9] also hinders the integration to current interconnect technologies. The criteria for evaluating a new barrier/liner are listed in Fig. 1d. Not until very recently, has a large-area, transfer-free, and BEOL-compatible MoS$_2$ barrier been realized[13].

In this work, we provide an approach that converts Ta, the current industry liner material, to a 2D layered material, namely tantalum sulfide ($TaS_x$), to serve as an ultra-thin barrier and a liner at the same time. The ~1.5 nm thick polycrystalline $TaS_x$ film is obtained on dielectrics (to demonstrate it can be deposited on IMDs) at a BEOL-compatible temperature. Our studies mainly focus on the comparison between (i) Ta and $TaS_x$ for liner properties and (ii) TaN and $TaS_x$ for diffusion barrier properties. Note, this is the first time that adhesion properties of 2D layered materials are thoroughly explored for the liner application. The results show that $TaS_x$ possesses superior liner and barrier properties simultaneously. In contrast, conventional TaN/Ta bilayer requires two types of thicker films to meet the requirements for both barrier and liner. By replacing the TaN/Ta bilayer with a 2D $TaS_x$ layer, the percentage of Cu volume in an ultra-scaled interconnect can be significantly increased to achieve low line resistance. Showing the promising barrier/liner properties of $TaS_x$, our work may encourage future development of high quality, single-layer tantalum sulfide films for BEOL applications.

**Results and Discussions**

**Material Preparation and Characterization.** 2D materials are normally grown at high temperatures (> 800 °C), which does not satisfy the requirement of BEOL-compatibility. Therefore, in this work, plasma-enhanced-chemical-vapor-deposition (PECVD)[16–18] was chosen to lower the growth temperature by utilizing the energy of the remote plasma. Centimeter-scale, uniform $TaS_x$ was prepared by converting a polycrystalline Ta film in a hot-zone furnace illustrated in Fig. 2a. The Ta film was pre-deposited on a Si/$SiO_2$ substrate by an e-beam evaporator. Hydrogen sulfide ($H_2S$) was used as the precursor while Ar was used as the carrier gas. The flow rates for both gases were 10 sccm. The growth temperature, time, and the plasma power were set at 400 °C, 20 minutes, and 70-80 W, respectively. During the growth, the remote plasma dissociated $H_2S$ into H and S radicals. S radicals reacted with Ta and converted Ta to $TaS_x$. The mechanism of conversion is illustrated in Fig. 2b, by using 1T-tantalum disulfide ($TaS_2$) as an example. This method was modified from the method for low-temperature $MoS_2$ growth[17,18]. This growth/conversion process produced a $TaS_x$ film. Detailed material analyses will be discussed in the following paragraphs.

In addition to the BEOL-compatible growth temperature, preparation of large-area films is also crucial to the integration with the interconnect technology. However, some of the 2D material synthesis recipes can only produce 2D flakes with micrometer-scale areas. Figure 2c confirms a large-scale, uniform polycrystalline Ta to $TaS_x$ conversion was realized using our PECVD method.

As a demonstration, a ~3 nm Ta film was deposited on $SiO_2$(90 nm) grown on a Si(100) surface. After the conversion process through PECVD, an ~8 nm $TaS_x$ film was formed at growth temperature of either 400 °C or 800 °C, as confirmed by the cross-section transmission electron microscopy (TEM) image in Fig. 2d. The increased film thickness is consistent with the previously reported $MoS_2$ conversion[17]. The different colors between 400 °C and 800 °C conversion shown in the optical microscope images of Fig. 2c may imply different material properties, as will be discussed further. A layered structure can be observed in the TEM images. Some oxidized $TaS_x$ can be observed on the top of the film from the TEM image due to surface oxidation). Unreacted Ta can also be found at the bottom (from TEM) of the $TaS_x$ film grown at 400 °C, which can be attributed to the insufficient energy at the lower process temperature. These thick (~8 nm) $TaS_x$ films were intentionally selected for better imaging of the layered structures. For all electrical tests that will be discussed, $TaS_x$ with the thickness of interest (~1.5 nm) will be used.

Raman spectra in Fig. 2e reveal that the 800 °C $TaS_x$ is $1T\text{-}TaS_2$[19,20], as the related peak at around 100 $cm^{-1}$ was observed. The peak can only be resolved into multiple $TaS_2$ peaks at low temperatures[19]. However, this peak was not observed in the 400 °C $TaS_x$. The material was further investigated by X–ray photoelectron spectroscopy (XPS). Figure 2f shows the Ta 4f and S 2p core level spectra of the 400 °C $TaS_x$ film obtained by XPS using a monochromatic Al Kα radiation ($hv$=1486.7 eV). Spectral features associated with Ta-S bond are detected at 23.5 eV and 161.3 eV in the Ta $4f_{7/2}$ and S $2p_{3/2}$ core levels, respectively. Based on XPS analyses, the S:Ta ratio was estimated to be ~2.5 ($TaS_{2.5}$). An additional chemical state is detected at higher binding energy in the Ta 4f (26.2 eV) and S 2p (162.5 eV) core levels aside from that of the Ta-S bond. The binding energy of the new chemical state is lower than the reported value[20,21] for $Ta_2O_5$, but higher than the Ta-S chemical state, which is assigned to the formation of Ta-S-O compound with a stoichiometry of $TaO_{1.9}S_{2.1}$. The Ta 4f core level shows also a chemical state at lower binding energy (22.9 eV). The binding energy of this chemical state in the Ta 4f core level, in conjunction with the additional chemical state detected in the N 1s core level at 396.4 eV, suggests the formation of a covalent Ta-N bond (labeled in orange)[22]. The intensity ratio of Ta-N to Ta-S features is 0.14. We presume that Ta-N does not play a significant role in blocking Cu diffusion, based on its relative concentration and the non-stoichiometry ($TaN_{1.3}$). It is also important to note that the Ta 4f core level region is convoluted with O 2s core level at 24.0 eV. Based on the physical characterization, the composition of the converted film appears to consist of $TaS_{2.5}$ layers mixed with naturally-oxidized $TaS_{1.9}O_{2.1}$ region. It is noted that not only $TaS_2$ is well-known to exhibit layered structures, but $TaS_3$ has also been reported to exhibit such structures[23].

In addition, the phase diagram[24] shows that $TaS_3$ tends to be formed at lower temperatures with $H_2S$ as the precursor, while $TaS_2$ can be formed at higher temperatures. Based on these observations, we will denote this film as $TaS_x$ to avoid any misinterpretation. The main purpose of this paper is to study the liner and diffusion barrier properties of this "$TaS_x$" film.

Although the $TaS_x$ film shown in the cross-section TEM image (Fig. 2d) was around 8 nm, for the electrical test structures for the barrier and liner properties, $TaS_x$ thin films with thickness of the interest were adopted. The thickness is ~1.5 nm, close to the thickness of a two-layer (2L) $TaS_2$, as confirmed by atomic-force microscope (AFM) in Fig. 3a. Three-layer (3L) regions were also found occasionally, with the thickness of ~2 nm. A more precise thickness control can be achieved by using other Ta deposition methods. For the simplicity, this film will be denoted as "1.5 nm $TaS_x$" throughout the rest of the paper.

**Tests of Liner and Diffusion Barrier Properties.**

*Liner Properties*: Ta has been used as the "liner layer" to provide good Cu adhesion enabling survival after multiple damascene process steps[1–3]. However, it is also well known that the inelastic scattering at the Ta/Cu interface[25,26] can increase Cu resistivity, especially when Cu wire dimensions are extremely scaled. Since the reason of utilizing an ultrathin 2D layered $TaS_x$ barrier/liner is to address the issue of conductivity degradation in extremely-scaled interconnects, it is important to understand the impact of $TaS_x$ on the surface scattering of Cu interconnects. Here, such surface scattering at the $SiO_2$/Cu, Ta/Cu, and $TaS_x$/Cu interfaces was studied. To facilitate the analysis, ultra-thin Cu films with thicknesses of ~15 nm were deposited on the above-mentioned three different surfaces to enhance the contribution from the interface. Cu thin films were patterned into Kelvin structures for accurate resistance measurements, as illustrated in Fig. 3c. Details of the fabrication are described in the "Method" section. Figure 3d shows the measured Cu resistivity of multiple devices on the three surfaces. Despite the fact that Cu resistivity increases as its dimension decreases[27], the thinnest Cu (13 nm) on $TaS_x$ has the lowest resistivity among the three groups, indicating a more specular/elastic interface scattering. The enhancement of Cu conductivity by inserting/capping with a 2D layered material has also been observed and studied in other works[14,15,28,29]. It is believed that the inferior conductivity of Ta/Cu and $SiO_2$/Cu can be attributed to the perturbing localized interfacial states[15,28]. In contrast, the weaker interaction between 2D layered materials and Cu[30] is expected to result in a less perturbed interface and hence preserve the pristine Cu surface states more effectively[28].

In addition to the surface scattering effect of a liner on Cu resistivity, the wettability and adhesion of Cu to liners are also of great importance. The former is essential to provide a good surface for the Cu seeding layer and for the subsequent electroplating of Cu, while the latter is crucial for Cu to survive chemical-mechanical polishing (CMP) processes and directly impacts the electromigration lifetime of Cu wires[31]. To investigate the wettability, much thinner Cu films (~10 nm) were deposited on different surfaces. The left panel of Fig. 3e reveals lots of cracks on the Cu film deposited on $SiO_2$, indicating a poor wettability. In contrast, $TaS_x$ provides a good wettability as Ta does, as can be observed in the middle and right panels of Fig. 3e. The adhesion of Cu on $TaS_x$ was verified by a tape test method[32]. After the preparation of $TaS_x$, ~80 nm Cu was deposited on $TaS_x$ followed by the attachment of a 3M Scotch® Tape for a simple adhesion test[32]. After detaching the tape, only the regions with $TaS_x$ had Cu left; Cu on $SiO_2$ regions was detached by the tape, as shown in Fig. 3f. All the test results above indicate that $TaS_x$ can serve as a favorable liner for Cu.

***Diffusion Barrier Properties***: The diffusion barrier properties of $TaS_x$ were tested by time-dependent dielectric breakdown (TDDB) measurements[33–36]. TDDB measurement is a standard test method to evaluate gate dielectric reliability as well as Cu diffusion in the dielectrics (IMDs) of interconnects. Here, a capacitor structure[34] was adopted for TDDB measurements to study the intrinsic diffusion barrier properties of $TaS_x$. Extrinsic breakdown[34,37] due to the CMP process can occur in interconnect damascene structures, which affects a sensitive evaluation of $TaS_x$ for the barrier application. In the measurement, a constant electric field was applied across a capacitor structure to drive Cu ions into the dielectric, as depicted in the inset of Fig. 4a. Note that the mass transport of Cu ions driven by the electric field is actually "Cu drift". Nevertheless, we still denote it as "Cu diffusion" because of the convention terminology used in the field. The Cu ions driven into the dielectric can cause Cu-induced breakdown by forming conductive paths and/or by lowering the barrier of Poole-Frenkel conduction[35]. A superior diffusion barrier is expected to mitigate Cu diffusion, which prolongs the time-to-breakdown ($t_{BD}$) of the devices. Figure 4a shows the current evolution with the stress time at 7 MV/cm of multiple devices. Sudden current jumps indicate device breakdown. It can be observed that $t_{BD}$ of most devices increases when Ta is converted to $TaS_x$, suggesting improved diffusion barrier properties. The values of $t_{BD}$ of multiple devices were further plotted in the statistical distribution shown in Fig. 4b, where each point was obtained from one device. At a certain electric field, the device with the shorter/longer $t_{BD}$ was assigned to have the lower/higher value of the cumulative probability. Therefore, a straight line with a positive slope can be obtained. Medium-time-to-failure (TTF$_{50\%}$) was extracted from the figure for an easier comparison between different materials. From Fig. 4b, it is obvious that TaS$_x$ devices have ~6 times longer lifetime than Ta devices at 7 MV/cm.

Although TaN mainly serves the barrier function in the conventional barrier/linear bilayer, Ta can also contribute to the blocking of Cu diffusion. Figure 4c indicates that 1.5 nm TaS$_x$ has similar barrier properties as 3 nm Ta, which is essential to the barrier/liner thickness scaling requirement. Since TaN plays the dominant role in mitigating Cu diffusion, TaS$_x$ is also benchmarked to TaN results from other work[8]. It can be observed in Fig. 4d that 1.5 nm TaS$_x$ has similar performance as 2 nm TaN in terms of mitigating Cu diffusion. Based on the experimental results of liner and diffusion barrier properties, it is possible to reduce a 5 nm TaN/Ta stack (2 nm TaN + 3 nm Ta) to a 3 nm TaS$_x$ layer while maintaining or even improving the barrier/liner performance. Further scaling can be achieved given the 2D material nature of TaS$_x$. The projection of significant resistance reduction benefitting from the maximized Cu volume is provided in Fig. 4e.

It is well know that the columnar grain-boundaries (GB) structures in TaN/Ta provide paths for faster Cu diffusion[38–40]. It was also predicted that GBs should also be the dominant diffusion paths for Cu diffusion through 2D layered materials[41]. To better understand the Cu diffusion behavior through these new materials, we have performed density-functional-theory (DFT) calculations using TaS$_2$ as the material for analysis. The computation details can be found in the Supplementary Information. The energy barriers for Cu diffusion through GBs and through the intralayer were both calculated. The two types of diffusion were illustrated in Fig. 5a. Various TaS$_2$ GB structures were compared[42–44], as illustrated in Fig. S1, to find the lowest energy configuration. (Fig. S2). It is discovered that the energy of Cu adsorbed at the GB shown in Fig. 5b is ~1.5 eV lower than that at the interlayer site (The adsorption energies at the GB core and at the sites far from the GB core were calculated to be -4.154 eV and around -2.6 eV, respectively). This suggests that the energy barrier for Cu diffusion from the GB to the interlayer site should be at least 1.5 eV, which is higher than that of Ta/TaN[38,39]. We also find that the intralayer diffusion (Fig. 5c) is very fast, with an energy barrier of only 0.25 eV, as shown in Fig 5d. These results summarized in Fig. 5e suggest that the enhanced barrier performance of TaS$_x$ compared with conventional TaN/Ta is likely to arise from the difference in GB, which blocks the Cu diffusion more effectively. Further improvement of the TaS$_x$ grain size can lead to even better diffusion barrier properties[9].

**Conclusion**

In summary, a ~1.5 nm TaS$_x$ barrier/liner is realized by converting Ta with PECVD at a BOEL-compatible temperature, providing a near-term solution for industry to improve the reliability/performance of current interconnect technology. This work considers critical integration aspects of this 2D material in terms of liner and barrier properties. The liner properties are evaluated based on (i) surface scattering at various interfaces, (ii) wettability that TaS$_x$ can provide for the subsequent Cu seeding layers, and (iii) adhesion of Cu to TaS$_x$. The TaS$_x$ film passes all the tests, which is demonstrated in 2D layered materials for first time. The diffusion barrier properties are analyzed by time-dependent dielectric breakdown (TDDB) measurements. The TDDB results show better barrier properties after Ta being converted to TaS$_x$. Additionally, the benchmark of TaS$_x$ to TaN indicates enhanced Cu blocking capability. The test results of liner and diffusion barrier properties are summarized in Table I. Our accomplishments compared to other works are summarized in Table II. Further improvement of the film quality can be expected to bring even better barrier performance. Based on the evaluations, a conventional TaN/Ta bilayer stack can be replaced by an ultra-thin TaS$_x$ layer to maximize the Cu volume for ultra-scaled interconnects. Further development of CVD or ALD based growth methods may realize even thinner (single-layer; < 1 nm) and more uniform 2D TaS$_x$ barrier/liner.

## Methods

**Preparation of TaS$_x$ film**. Ta was first deposited by an e-beam evaporator on Si/SiO$_2$ at a rate of ~0.5 Å/s. After loading the Ta sample to the PECVD system, the tube furnace was pumped down. Nitrogen purging/chamber pumping were conducted repeatedly for ten times to remove moisture and other possible contamination sources from the ambient during the sample loading step. Meanwhile, the temperature was set at 400 °C, which was reached after 5-10 minutes. After reaching the base pressure of ~10 mTorr, Ar with a flow rate of 10 sccm was introduced. The pressure at this time was ~180 mTorr. Then, H$_2$S with a flow rate of 10 sccm was introduced and the pressure reached ~320 mTorr. After waiting for 5 minutes to stablize the temperature and flow rate, plasma power was turned on and slowly increased to 70 W. At this power, the pressure changed to ~420 mTorr. After 20 minutes of growth, the plasma was turned off, followed by turning off the gases and the furnace heater.

**XPS analysis.** Monochromatic XPS was employed in this work using a system described elsewhere[45]. The *AAnalyzer* software was used for XPS peak analysis[46]. Active Shirley background and Voigt line shape was employed for peak fitting. For all chemical states detected at Ta 4$f$ core level, the binding energy separation between 4$f_{7/2}$ and 4$f_{5/2}$ is 1.9 eV with the same Lorentzian (0.11). Gaussian is larger for Ta-O-S chemical state due to formation of new bonds. Similar for S $2p$ core level, binding energy separation (1.1 eV) and Lorentzian (0.11) for $2p_{3/2}$ and $2p_{1/2}$ is the same for S-Ta and S-Ta-O chemical states. While Gaussian is larger and broader for S-Ta-O bonding. Corresponding to Ta-S-O chemical state at S $2p$ and Ta $4f$ core levels additional chemical state observed at O $1s$ core level at lower binding energy (~530 eV).

**TEM analysis**. STEM cross-sectional samples were prepared with a FEI Nova 200 dual-beam FIB/SEM by using the lift-out method. The region of interest above the Al metal pad was protected during the focused ion beam milling, by depositing $SiO_2$ and Pt layers on top of the sample. Both high resolution transmission electron microscopy (HREM) images, atomic STEM HAADF and bright field (BF) images were obtained in a JEOL ARM200F microscope equipped with a spherical aberration (Cs) corrector (CEOS GmbH, Heidelberg, Germany) and operated at 200 kV. The corrector was carefully tuned by the Zemlin-tableau method with Cs = 0.5 µm and the resolution was demonstrated to be around 1 Å.

**Fabrication of MOS capacitor and Kelvin structures**. Heavily-doped Si (resistivity < 5 mΩ-cm) substrates with 90 nm $SiO_2$ were used for all sample fabrication. Ta/Cu was always deposited *in-situ*, while $TaS_x$/Cu encounter air exposure after the Ta deposition and $TaS_x$ conversion. For the fabrication of capacitor structures, on top of Ta or $TaS_x$, Cu/Al (~30 nm/20 nm) electrodes with diameters of 100 µm were deposited by e-beam evaporation using a shadow mask, with Cu in contact with the 2D material and Al on the very top as a passivation layer. The sample was then coated with photoresist and placed into 6:1 buffered oxide etch (BOE) to etch away the $SiO_2$ on the back side of the Si substrate, followed by 50 nm Al deposition to form an ohmic contact to the Si substrate. Finally, the top photoresist was removed by acetone. For the Kelvin structure, the same $Si/SiO_2$ substrate with Ta or $TaS_x$ on top (or $SiO_2$ only) was first coated with poly (methyl methacrylate) (PMMA). 30 minutes of forming gas annealing was used to remove residues on $TaS_x$ from the growth before PMMA coating. E-beam lithography was then used to write the desired patterns, followed by a development in IPA: water = 3: 1. Cu was e-beam-evaporated on the sample. Finally, a lift-off process was used to form the desired Cu patterns.

**DFT calculations**. DFT calculations were performed using the Vienna Ab-initio Simulation Package (VASP)[47,48] with Projector Augmented Wave (PAW) pseudopotential[49] and Perdew-Burke-Ernzerhosf (PBE) exchange-correlation functional[50]. A kinetic energy cut-off of 400 eV was used for the plane-wave expansion, and all atomic positions were fully relaxed until the final force on each atom was less than 0.01 eV/Å. The vdW interaction in layered systems was incorporated using Grimme's method[51]. To model the interlayer diffusion, we used a 4x4 supercell of bilayer $TaS_2$. The structures of GBs and their energies can be found in the SI.

# Acknowledgments

This work was supported in part by NEWLIMITS, a center in nCORE, a Semiconductor Research Corporation (SRC) program sponsored by NIST through award number 70NANB17H041. CL and ZC also acknowledge financial support from NSF (Grant No. CCF-1619062). The DFT calculations are supported by Welch Foundation (Grant No. F-1959-20180324) and the startup grant from The University of Texas at Austin (UT Austin). The calculations used computational resources located at the National Renewable Energy Laboratory (NREL) sponsored by the DOE's Office of Energy Efficiency and Renewable Energy (EERE), and used the Texas Advanced Computing Center (TACC) at UT Austin.

**Author Contributions**

Z.C. conceived and managed the project. C.L. discovered the novel $TaS_x$ barrier/liner material and designed and performed all the device fabrications and electrical measurements. M.C. and L.W. conducted all the STEM analyses. D.Z. conducted the XPS measurements while A.K. and R.A analyzed the XPS data. R.W., M.K., and Z.C. supervised all experiments. W.G. and Y.J. performed DFT calculations under the guidance of Y.L. All authors took part in discussion on results and preparation of the manuscript.

**Competing interests**: The authors declare no competing financial interests.

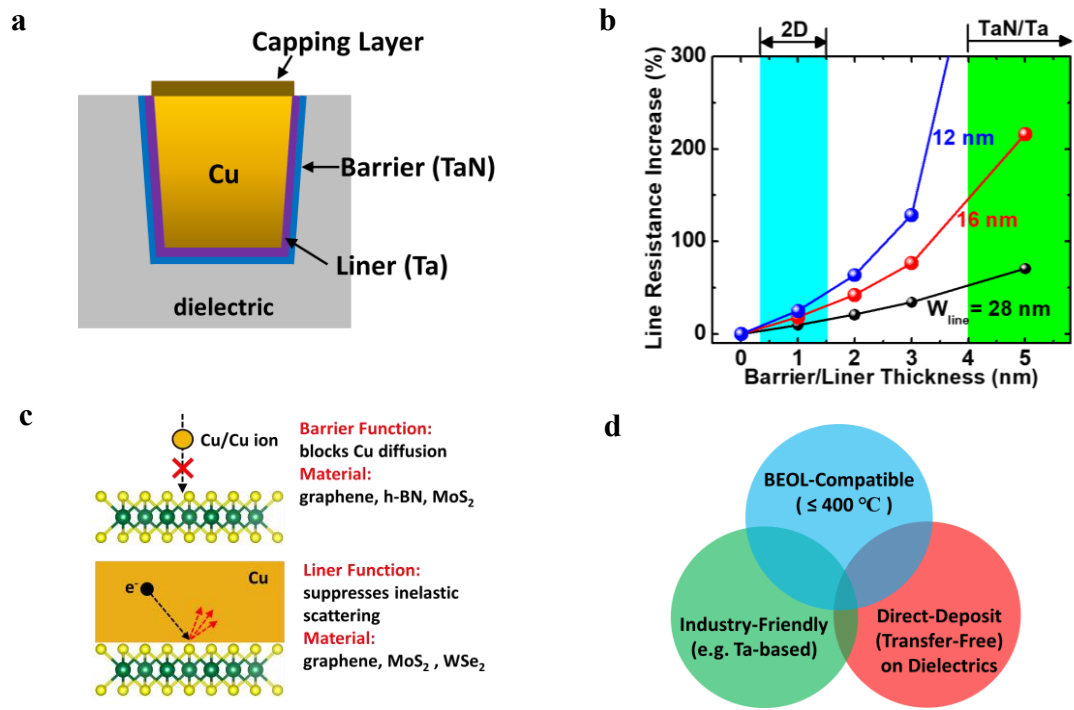

**Fig. 1** Advantages of using 2D layered materials as Cu diffusion barrier and liner. **a** Schematic of the cross-section of a Cu interconnect. A resistive barrier/liner bilayer surrounds Cu and occupies a certain portion of the interconnect. This structure is also known as the damascene structure. **b** Effects of barrier/liner thickness on interconnect resistance. Only ultra-thin 2D layered materials can prevent significant resistance increase, especially for ultra-scaled interconnects. **c** Recent discoveries of the benefits that 2D layered materials can bring to Cu interconnects. **d** Requirements of a new barrier/liner material to replace conventional TaN/Ta bilayer. The BEOL-compatible TaS$_x$ directly deposited on dielectrics in this work meets all the requirements.

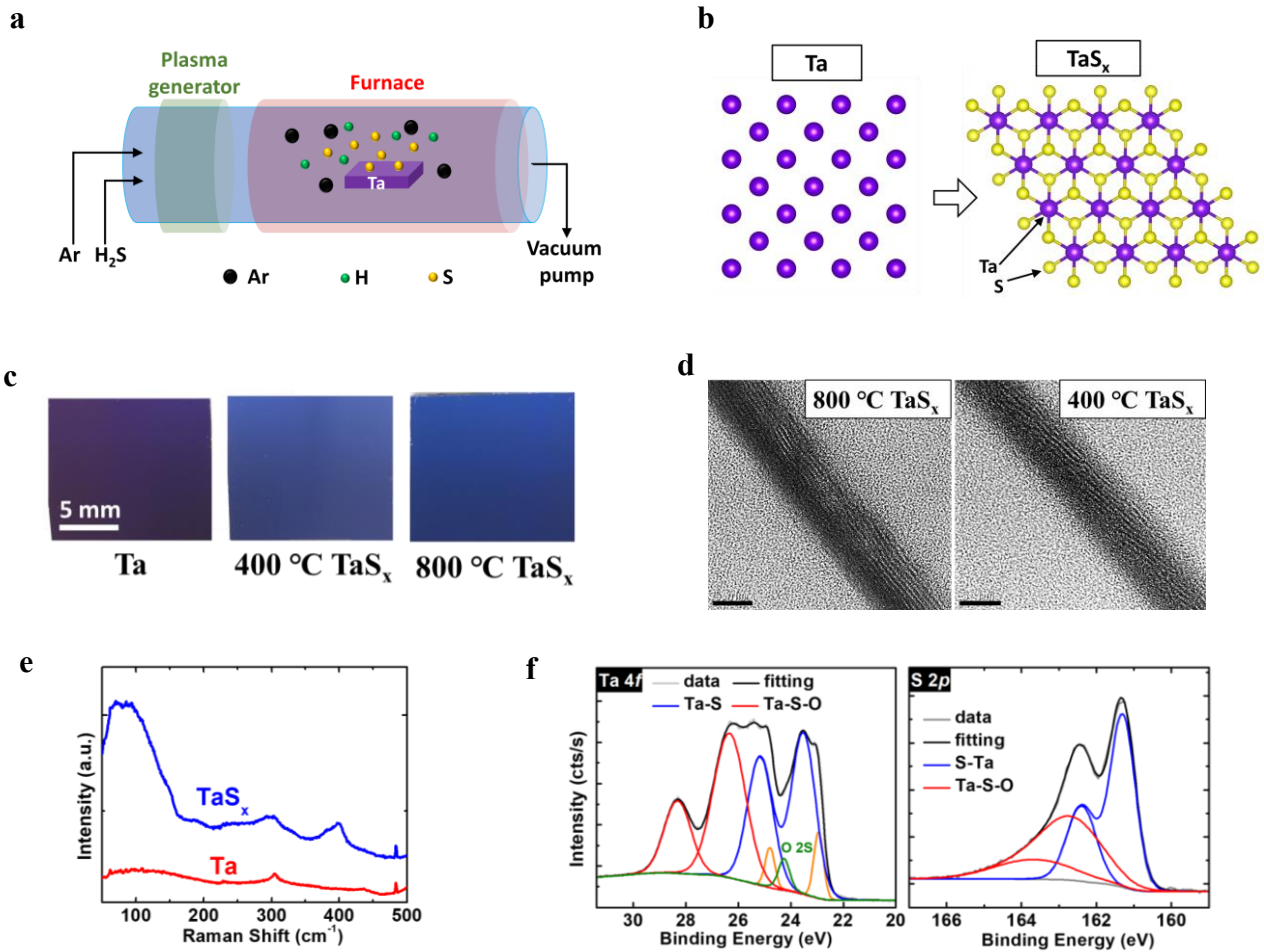

**Fig. 2** Material preparation and characterization of tantalum sulfide. **a** Schematic of the PECVD system for $TaS_x$ growth. Plasma assists in dissociating $H_2S$ and hence reduces the growth temperature. Sulfur radicals react with Ta and convert Ta to $TaS_x$. **b** Illustration of the conversion from Ta to $TaS_x$. The structure of $1T\text{-}TaS_2$ is adopted for simplicity. **c** Top-view of large-area, continuous $TaS_x$ converted from Ta. **d** TEM images showing the layer structures of thicker (~8 nm) $TaS_x$ grown at different temperatures. **e** Raman spectra of $TaS_x$ on $Si/SiO_2$ (90 nm). Characteristic peaks of $1T\text{-}TaS_2$ is identified in 800 °C $TaS_x$. The wavelength of the laser used for Raman measurements was 532 nm. **f** XPS analysis of 400 °C $TaS_x$. Ta-S, Ta-S-O, and some Ta-N bonds (in orange) are detected. The ratio of Ta to S is 1: 2.5. Oxidized Ta is unavoidable due to the air-exposure before the measurement. The small amount and non-stoichiometric Ta-N cannot block Cu diffusion. An optimized $TaS_x$ can bring even better diffusion barrier properties, while the $TaS_x$ film demonstrated here has already shown to be able to block Cu diffusion efficiently. The scale bars are 5 mm and 5 nm in **c** and **d**, respectively.

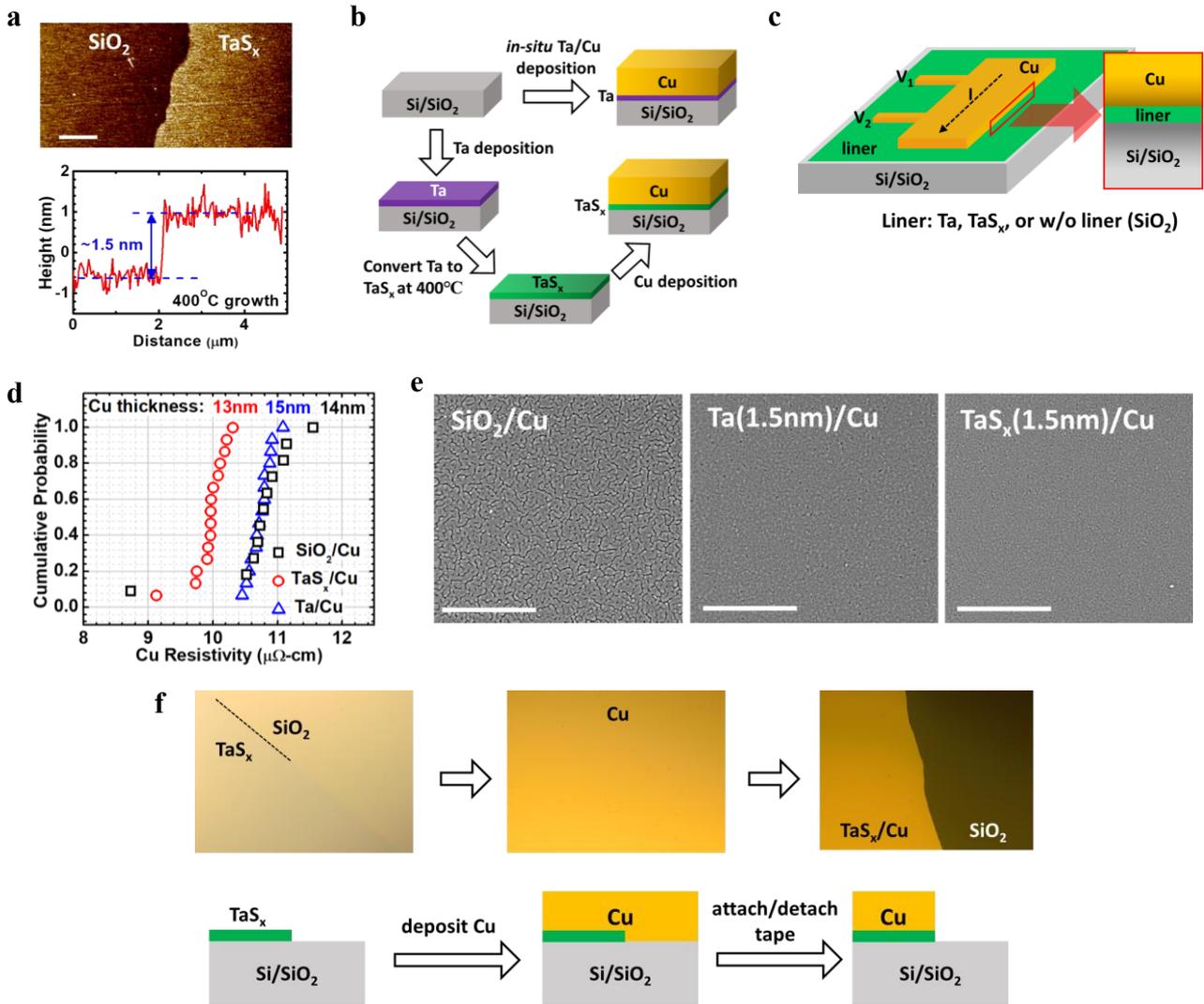

**Fig. 3** Device structures and tests of liner properties of 400 °C $TaS_x$. **a** AFM profile of 1.5 nm $TaS_x$ grown at 400 °C. This film is used for all the liner and barrier tests. **b** Sample fabrication procedure. Ta/Cu stack is deposited *in-situ*, while $TaS_x$ has been exposed to air before Cu deposition. An *in-situ* Cu deposition on $TaS_x$ could bring an even better performance. **c** Cu thin film patterned into Kelvin structure for resistance measurements. Ultra-thin Cu (~15 nm) film is adopted to enhance the contribution of the interface. **d** Cu resistivity on various surfaces. In general, thinner Cu is expected to have a higher resistivity. Nevertheless, the thinnest Cu (13 nm) on $TaS_x$ has the lowest resistivity, indicating suppression of surface scattering at the $TaS_x$/Cu interface. **e** Wetting properties of Cu tested by depositing ultra-thin Cu (~10 nm) on different surfaces. Numbers of cracks can be observed when Cu is on $SiO_2$, while Cu on 1.5 nm Ta and 1.5 nm $TaS_x$ have smooth morphologies. The results show that $TaS_x$ can provide a good surface as Ta does for Cu seeding layers, which is important for the subsequent Cu electroplating **f** Adhesion tests using the tape method. ~80 nm Cu is deposited on $TaS_x$ or $SiO_2$. After detaching the tape, only Cu on $TaS_x$ remains, indicating that $TaS_x$ is a good liner for Cu to survive CMP processes. The scale bars are 1μm and 500nm in **a** and **e**, respectively.

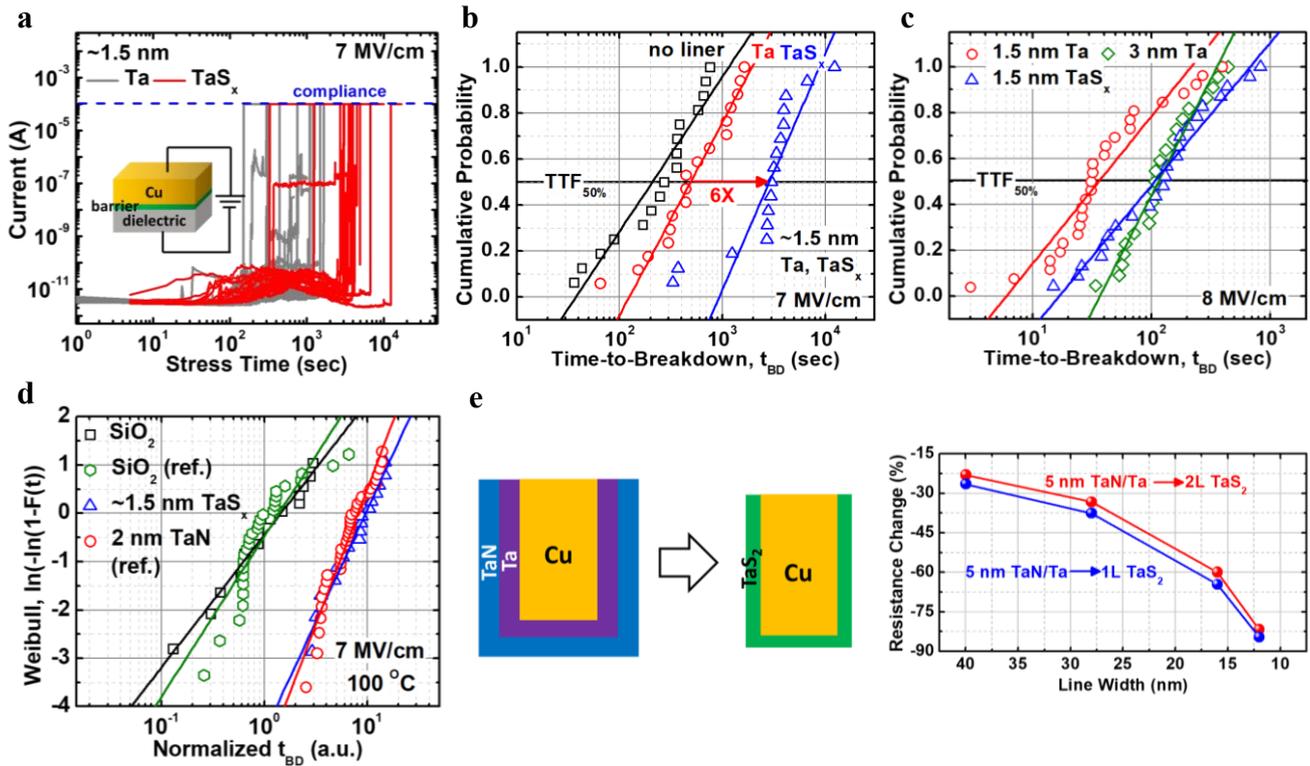

**Fig. 4** Tests of diffusion barrier properties of 400 °C TaS$_x$. **a** Current versus stress time of devices with Ta and TaS$_x$. Devices with TaS$_x$ have longer lifetime in general. **b** Statistical distribution of devices with Ta and TaS$_x$ barriers. By converting Ta to TaS$_x$, medium-time-to-failure (TTF$_{50\%}$) increases by 6 times. **c** 1.5 nm TaS$_x$ has similar diffusion barrier properties as a 3nm Ta. Since Ta also plays a role (minor) in blocking Cu diffusion, this result indicates the Ta liner thickness can be reduced by using a thinner TaS$_x$ and Cu volume can be increased. **d** Benchmarking diffusion barrier properties of ~1.5-nm TaS$_x$ against 2-nm TaN. Two materials have a comparable diffusion barrier properties. Improvement of TaS$_x$ qualify can further reduce the required thickness. **e** Resistance reduction at various line widths with the thinner barrier/liner layer by using TaS$_x$. The improvement in narrower interconnect is more significant. A method to have a single-layer/high-quality TaS$_x$ can lower the resistance much more tremendously.

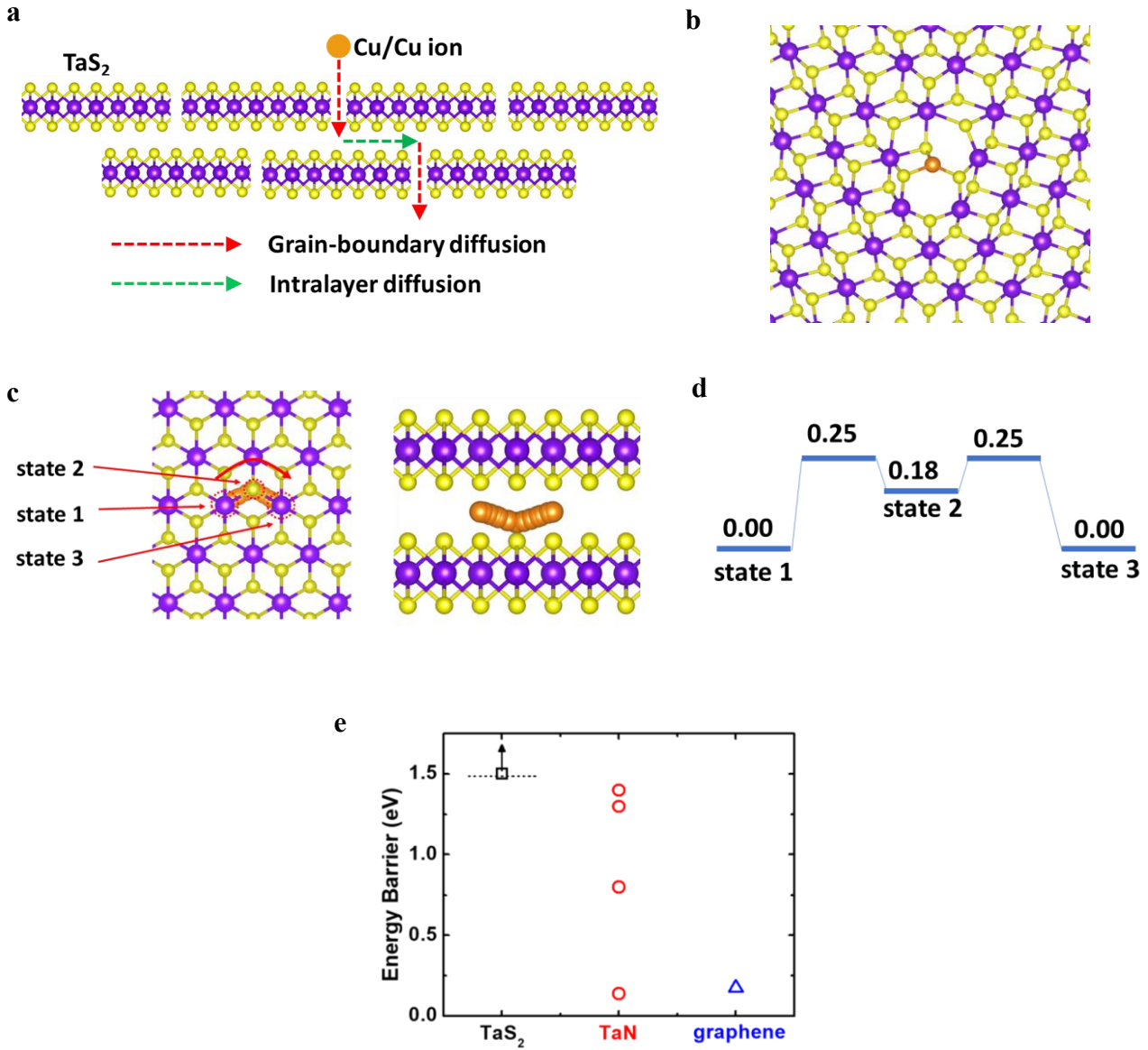

**Fig. 5** DFT calculations of Cu diffusion through a 2D TaS$_2$ barrier. **a** Illustration of grain-boundary (GB) and interlayer diffusion. **b** Atomic structure of Cu bound at GB of TaS$_2$. **c** Atomic pathway of intralayer diffusion of Cu in layered TaS$_2$. **d** Energies of the states along the diffusion pathway. **e** Comparison of the energy barriers for Cu diffusion through GBs of various materials. The values for TaN and graphene are from other works[38,39,41]. The calculation indicates TaS$_2$ has a sufficient energy barrier of at least 1.5eV to block Cu diffusion.

**Table I.** Test results of liner and diffusion barrier properties. The liner properties are tested in three aspects, while the analysis of diffusion barrier property focuses on the capability of blocking Cu diffusion.

| **Liner Properties** | | **Diffusion Barrier Property** | |
|---|---|---|---|
| Cu surface scattering | TaS$_x$ reduces surface scattering →Cu resistivity decreases | compared to Ta | better |
| Cu wetting | no obvious cracks in Cu thin films deposited on TaS$_x$ →TaS$_x$ provides a good surface for Cu seeding layer | compared to TaN | similar |
| Cu adhesion | pass tape test →could survive CMP process | | |

**Table II.** Comparison of different works using 2D layered materials as Cu diffusion barriers. Only the TaS$_x$ barrier/liner in this work satisfies all of the requirements.

| Material | Growth Temperature (BEOL-Compatible?) | Transfer-Free? | BARRIER PROPERTY? | Liner Property? |
|---|---|---|---|---|
| Graphene[8] | 1000 ℃ (NO) | NO | YES | NO |
| Graphene/ graphene oxide[7] | 750 ℃ (NO) | NO | YES | NO |
| Graphene[5,6] | NO | NO | YES | NO |
| Graphene[10] | 550 ℃ (NO) | YES | YES | NO |
| Graphene[12] | 400 ℃ (YES) | YES | YES | NO |
| MoS$_2$[11] | 850 ℃ (NO) | YES | YES | NO |
| TaS$_x$ (This Work) | 400 ℃ (YES) | YES | YES | YES |

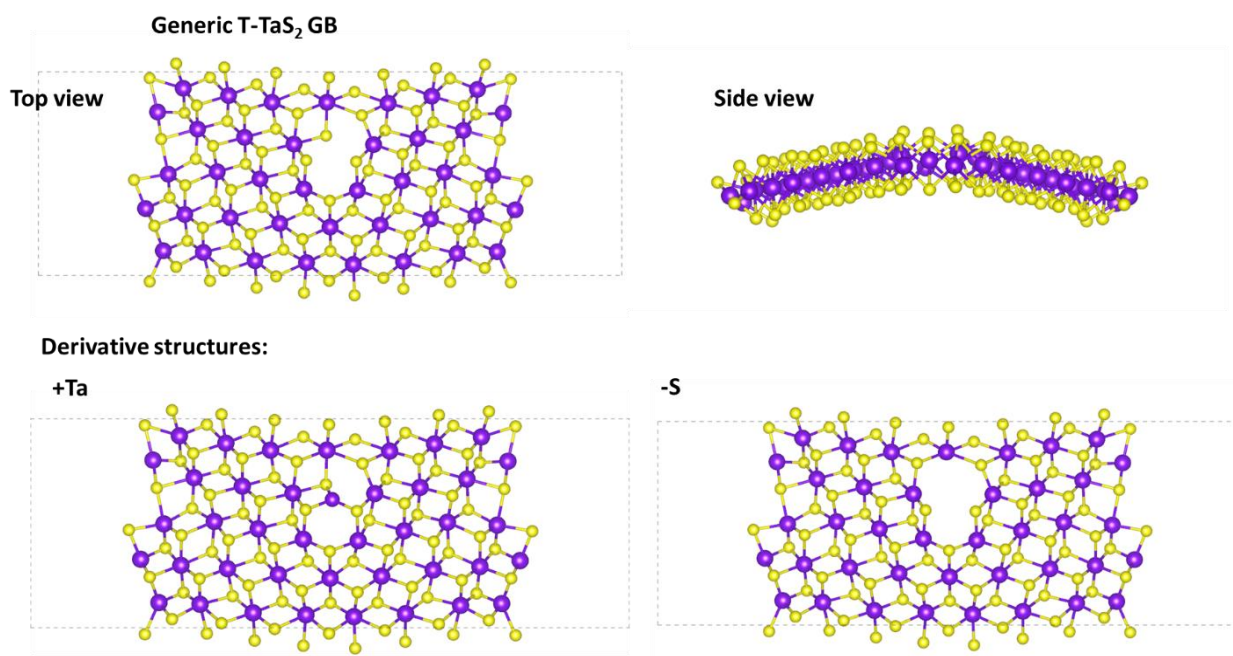

**Fig. S1**. Three types of possible GB structures. The "generic" structure has the Ta:S ratio same as that in perfect material (i.e. 1:2). +Ta means adding one more Ta atom into the GB core, and –S means removing one S atom from the GB core.

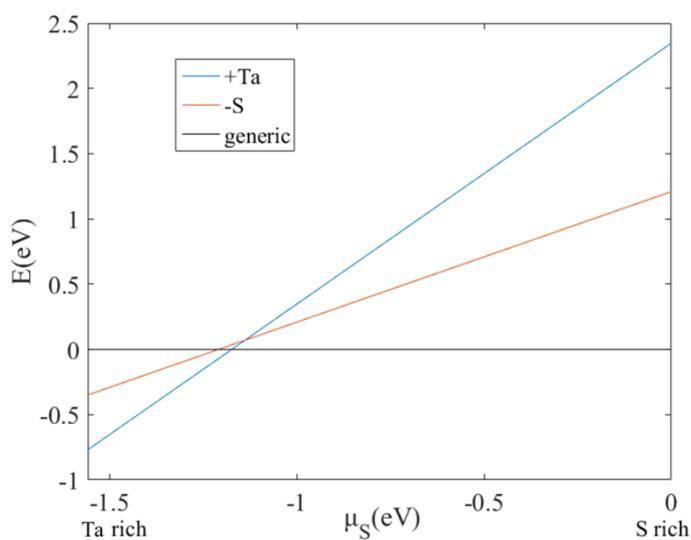

**Fig. S2**. Formation energies of the GBs shown in Fig. S1, as a function of the S chemical potential. The formation energy of generic GB is set to zero as reference. At Ta/S-rich condition, the chemical potential of Ta/S equals to the energy of bulk Ta/S. The plot shows that for most chemical potential, the generic GB structure is more energetically favorable, which is thus chosen to study Cu adsorption.